\documentclass[11pt,a4paper]{article}
\usepackage{microtype}

\usepackage{fullpage}

\usepackage{amsthm}
\usepackage{amsmath}
\usepackage{amssymb}
\usepackage{algorithmic}
\usepackage[linesnumbered, ruled, boxed]{algorithm2e}
\SetProcNameSty{textsc}

\SetKwBlock{MyBlock}{}{}
\SetKwProg{MyCaseOne}{Case 1}{ do}{end}

\usepackage{tikz}
\usetikzlibrary{shapes.geometric}
\usetikzlibrary{arrows}
\usetikzlibrary{decorations.pathreplacing}
\tikzset{
itria/.style={
  draw,dashed,shape border uses incircle,
  isosceles triangle,shape border rotate=90,yshift=-1.45cm},
>=stealth,
triangle/.style={isosceles triangle,draw,shape border rotate=90, dashed, minimum height=10mm, minimum width=15mm, inner sep=0},
dash/.style={edge from parent/.style={dashed,draw}},
}

\usepackage{todonotes}
\newtheorem{property}{Property}

\newcommand\esplit[2]{\ensuremath{\textsc{Split}(#1, #2)}}
\newcommand\emerge[2]{\ensuremath{\textsc{Merge}(#1, #2)}}
\newcommand\ejoin[2]{\ensuremath{\textsc{Join}(#1, #2)}}

\newcommand\emakeset[1]{\ensuremath{\textsc{MakeSet}(#1)}}

\newcommand\dsplit{\ensuremath{\textsc{Split}}}
\newcommand\dmerge{\ensuremath{\textsc{Merge}}}

\newcommand\dsearch{\ensuremath{\textsc{Search}}}
\newcommand\dshift{\ensuremath{\textsc{Shift}}}
\newcommand\dmakeset{\ensuremath{\textsc{MakeSet}}}
\newcommand\djoin{\ensuremath{\textsc{Join}}}

\newtheorem{lemma}{Lemma}

\newtheorem{theorem}{Theorem}

\title{Mergeable Dictionaries With Shifts}
\author{Philip Bille \and Mikko Berggren Ettienne \and Inge Li G{\o}rtz}

\begin{document} 
\maketitle

\begin{abstract}
    We revisit the mergeable dictionaries with shift problem, where the goal is to maintain a family of sets subject to search, split, merge, make-set, and shift operations. The search, split, and make-set operations are the usual well-known textbook operations. The merge operation merges two sets and the shift operation adds or subtracts an integer from all elements in a set. Note that unlike the join operation on standard balanced search tree structures, such as AVL trees or 2-4 trees, the merge operation has no restriction on the key space of the input sets and supports merging arbitrarily interleaved sets. This problem is a key component in searching Lempel-Ziv compressed texts, in the mergeable trees problem, and in the union-split-find problem.  

We present the first solution achieving O(log U) amortized time for all operations, where $\{1, 2, \ldots, U\}$ is the universe of the sets. This bound is optimal when the size of the universe is polynomially bounded by the sum of the sizes of the sets. Our solution is simple and based on a novel extension of biased search trees.

\end{abstract}

\section{Introduction}

We consider the \emph{mergeable dictionary with shifts} problem. 
A mergeable dictionary with shifts maintains a dynamic collection of sets
$\mathcal{G} = \{G_1, G_2, \ldots, G_m\}$ from a totally 
ordered universe $\{1, 2, \ldots, U\}$ subject to the following
operations (the sets $G_1, \ldots ,G_m$ need not be disjoint): 

\begin{itemize}
    \item $\textsc{Search}(G, j)$: Return the largest element from $G$ that is at most $j$ if any such element exists.
    \item $\textsc{Split}(G, j)$: Split $G$ into two sets
    $A = \{ x \in S \mid x \leq j\}$ and $B = \{ x \in S \mid x > j\}$,
    remove $G$ from $\mathcal{G}$ and insert $A$ and $B$.
    \item $\textsc{Merge}(A, B)$: Remove $A$ and $B$ from $\mathcal{G}$
    and insert $C = A \cup B$ instead.
    \item $\textsc{MakeSet}(j)$: Insert a new singleton set $G = \{ j \}$ in $\mathcal{G}$.
    \item $\textsc{Shift}(G, j)$: Shift all elements in $G$ by $j$, i.e.,
    $G = \{ g + j \mid g \in G\}$.
\end{itemize}
This problem is a key component in searching Lempel-Ziv compressed text \cite{Farach1998}, the mergeable trees problem~\cite{Georgiadis:2011:DSM:1921659.1921660}, and generalizations of the union-find-split problem~\cite{Lai2008}.

Standard binary search trees, e.g., AVL-trees or 2-4 trees, support $\textsc{Search}$ and $\textsc{Split}$ in logarithmic time, while $\textsc{Shift}$ and $\textsc{MakeSet}$ take constant time. Most standard binary search trees can also be extended to support the $\textsc{Join}$ operation that takes two sets where all the elements in one set are larger than the other and merge them into a single set. The $\textsc{Merge}$ operation has no such restriction on the input sets and supports merging arbitrarily interleaved sets. It is easy to show that sublinear worst-case bounds for $\textsc{Merge}$ are not possible. The $\textsc{Shift}$ operation is also straightforward to implement on most binary search trees but non-trivial in the combination with the $\textsc{Merge}$ operation. 

The first non-trivial bound for mergeable dictionaries with shifts was given by Farach and Thorup~\cite{Farach1998} who showed that a simple folklore merge strategy called \emph{segment merge} yields an $O(\lg U \lg n)$ amortized time for the operations where $n$ is the sum of the sizes of the sets. This solution uses standard binary search trees with logarithmic time $\textsc{Join}$ and $\textsc{Split}$ operations and constant time $\textsc{Shift}$. Lai~\cite{Lai2008} conjectured that this bound is  optimal, but this was disproven by Iacono and \"{O}zkan~\cite{Iacono2010}, who showed how to support all operations except $\textsc{Shift}$ in $O(\log U)$ amortized time\footnote{The bound is stated as $O(\log n)$ in the paper since they assume $U = n$.}. Iacono and \"{O}zkan claim that the $\textsc{Shift}$ operation can also be supported by their data structure within the same complexity, but give no proof. We believe that this is true, but implementing $\dshift{}$ operation efficiently in their framework is non-trivial,
in part because their solution requires sets to be disjoint. Furthermore, the implementation and analysis of their solution is quite involved and require 25+ pages in the full technical report. 



More recently, Karczmarz~\cite{karczmarz:LIPIcs:2016:6028} gave a very simple solution without the $\textsc{Shift}$ that achieves $O(\log U)$ amortized time. This solution is based on binary trie representations of sets combined with word-level parallelism. As the author mentions, this approach does not extend to easily support the $\textsc{Shift}$ operation. 
It does however handle infinite/dynamic universes.
Obtaining amortized logarithmic time complexity for the mergeable dictionary problem while supporting both infinite universes and the $\textsc{Shift}$ operation is still an open problem.

\subsection{Our Results}
We show the following main result. 
\begin{theorem}\label{thm:main}
    There exist a mergeable dictionary with shifts data structure supporting all operations in  $O(\lg U)$ amortized time.
    
    For a set $G$, let $U_G=\max(G)-\min(G)$.
    The \dsearch{} and \dsplit{} operations take
    $O(\lg U_G)$ worst-case and amortized time, and the \dmakeset{} and \dshift{}
    operations take $O(1)$ worst-case and amortized time. The amortized time of the \dmerge{} operation is $O(\lg U_G)$, where $G$ is the set output by the operation.
\end{theorem}
We note that the complexity of our mergeable dictionary operations 
only depends on the ``local universe'' of the sets involved in the operation. This implies that the $O(\log U_G)$ bounds holds even if the upper bound $U$ of the universe changes.

This is the first solution to the mergeable dictionary with shift problem using $O(\log U)$ amortized time (with an implementation and analysis of the $\textsc{Shift}$ operation). For universes bounded in size by a polynomial in the sum of the sizes of the input, the bound is optimal~\cite{Iacono2010}. Thm.~\ref{thm:main} improves the result by Farach and Thorup~\cite{Farach1998} by a logarithmic factor. We match the bound of Iacono and \"{O}zkan~\cite{Iacono2010} and Karczmarz~\cite{karczmarz:LIPIcs:2016:6028} but add support for the $\textsc{Shift}$ operation.

To obtain Thm.~\ref{thm:main} we design a modified version of the segment merge strategy carefully
designed to work with biased search trees. This leads to a surprisingly simple analysis relative
to previous work. In particular we avoid complicated finger operations and analysis.





\subsection{Outline}
In Section \ref{sec:segmentmerge} we explain the folklore merge strategy 
in combination with binary search trees described in Farach and Thorup~\cite{Farach1998} and review the proof from Farach and Thorup~\cite{Farach1998} that yields an amortized $O(\lg n \lg U)$ 
solution to the mergeable dictionary problem.
Section~\ref{sec:biasedtrees} revisits the biased search tree by Ben~et~al.~\cite{BentST85}
and Section~\ref{sec:datastructure} give the details our weighting scheme.
We then move on to describe and analyze our biased segment merge operation in Section~\ref{sec:merge}.
In Section~\ref{sec:other-operations} we describe how to support shifts and analyze the amortized complexity of the remaining operations. Finally, we show how to handle intersecting
sets in Section~\ref{sec:intersect}.

\section{Segment Merge}\label{sec:segmentmerge} In this section we explain the segment merge algorithm described in Farach and Thorup \cite{Farach1998}, which our biased segment merge is based on.
The merge operation merges two arbitrarily interleaved ordered sets
$A$ and $B$. Assume that $A \cap B =\emptyset$ (we show how to lift this assumption later). We first consider the case where $ \min(A) < \min(B)$
and $\max(A) < \max(B)$.

The segment merge algorithm merges the ordered sets $A$ and $B$
by partitioning the two sets into a minimal number of \emph{segments}
$\{ A_1, \ldots, A_k\}$  and $\{ B_1, \ldots, B_k\}$ such that $A_i \subseteq A$ and $B_i \subseteq B$ 
and $\max(A_i) < \min(B_i)$ and $\max(B_i) < \min(A_{i+1})$ which are then subsequently joined together.

Given a set data structure that supports \dsplit{} and \djoin{} the merge operation is then performed as follows:

Initially set $C = \emptyset$. For $i = 1, \ldots, k$ do:
\begin{itemize}
    \item Set $A_i, A \leftarrow \esplit{A}{\min(B)}$
    \item Set $B_i, B \leftarrow \esplit{B}{\min(A)}$
    \item Set $C_i \leftarrow \ejoin{A_i}{B_i}$
    \item Set $C \leftarrow \ejoin{C}{C_i}$
\end{itemize}

\noindent After this process it is clear that $C$ is the ordered set $A \cup B$.


Using standard search trees the \dsplit{} and \djoin{} operations can be implemented in $O(\lg n)$ worst-case time
where $n = \sum_{G \in \mathcal{G}} |G|$. Thus the total time for a segment merge is $O(k \lg n)$ which in the worst case is $O(n \lg n)$.
However, the amortized complexity of segment merge with standard binary trees is $O(\lg U\lg n)$ as shown
by Farach and Thorup \cite{Farach1998}. As a warm-up we revisit their analysis, since we will reuse some of their definitions in our analysis.

Let $x^-$ and $x^+$ denote the predecessor and successor of $x \in G$
and define the size of the \emph{left gap} of $x$ to be $g^-(x) = 1$ if $x = \min(G)$ and $g^-(x) = x - x^-$ otherwise.
Similarly, define the size of the \emph{right gap} of $x$ as $g^+(x) = 1$ if $x = \max(G)$ and $g^+(x) = x^+ - x$ otherwise.
Define the \emph{potential} of a set $G$ as $$\phi(G) = \sum_{x \in G} ( \lg g^+(x) + \lg g^-(x))$$
and define the potential of the mergeable dictionary with shifts as 
$$
    \Psi(\mathcal{G}) = \lg n \sum_{G \in \mathcal{G}} \phi(G) 
$$

Clearly, the potential is always non-negative.
Now consider the potential of the data structure before and after a merge operation $C \leftarrow \emerge{A}{B}$.
It is easy to see that the only gaps that increase are the left gap of the element $\min(B)$
and the right gap of the element $\max(A)$. Clearly, the increase is bounded by $\max(C) - \min(C) \leq U_C$
causing a potential increase of $O(\lg n \lg U_C)$.
Whenever we insert a segment of a set between to elements
$x$ and $y$ where $x < y$ in another set, we halve either $g^+(x)$ or $g^-(y)$
causing the potential to decrease by at least $\lg n$.
It follows that the potential decreases by $\Omega(k \lg n - \lg n \lg U_C)$ when there are $k$ segments and thus the amortized cost of the merge operation is $O(\lg n \lg U_C)$.
None of the operations \dmakeset{}, \dsplit{}, \dshift{} and \dsearch{} increase the potential of the data structure.

If $\min(B) < \min(A)$ we swap the arguments to merge. If $\max(A) > \max(B)$ we
execute the following operations $B', B'' \leftarrow \esplit{B}{\max(A)}$, $C' \leftarrow \emerge{A}{B'}$,
and finally because $\max(C') = \max(A) < \min(B'')$ we can produce $C$ by joining $C'$ and $B''$.
Now the assumption is true for the merge because $\max(A) < \max(B')$
and the extra split and join operations both take $O(\lg n)$ time.


This shows  that standard search trees solve the mergeable dictionary problem in $O(\lg n \lg U_C)$ amortized time. We now
 move on to explain how to improve this bound to $O(\lg U_C)$ using biased search trees instead of standard search trees.

\section{Biased Trees}\label{sec:biasedtrees}
In this section we revisit the biased 2,3-trees by Bent~et~al.~\cite{BentST85}.

A biased 2,3-tree stores a set of $n$ keys in the leaves of a tree where all internal vertices
have 2 or 3 children. If $x$ is the $i^{th}$ leaf in left to right order it stores the $i^{th}$ key when sorted in increasing order
and internal vertices store the maximal and minimal key of its leaf descendants.
The \emph{weight} of a vertex $x$ is denoted $w_x$.
Every leaf is assigned a weight and the weight of an internal vertex is the sum of the weights
of its leaf descendants. The weight of a tree $T$ denoted $W_T$ is the weight of its root.
The \emph{rank} of a vertex $x$ is denoted $r(x)$,
and $r(x) = \lfloor \lg w_x \rfloor$ if $x$ is a leaf,
whereas $r(x) = 1 + \max\{r(y) \mid y \text{ is a child of } x\}$ if $x$ is not a leaf.
The rank of a tree $T$ denoted $r(T)$ is the rank of its root.
Let $y$ be the child of $x$. Then $y$ is \emph{major}
if $r(y) = r(x) - 1$ and \emph{minor} if $r(y) < r(x) - 1$. 
A 2,3-tree is biased when any neighboring sibling of a minor vertex is a major leaf.
A biased 2,3-tree have the following properties:

\begin{lemma}[Lemma 1, Bent et al.~\cite{BentST85}]
    \label{lem:max-rank}
    For any vertex $x, 2^{r(x) - 1} \leq w_x$, and if $x$ is a leaf then $2^{r(x)} \leq w_x < 2^{r(x) + 1}$.
\end{lemma}

\begin{lemma}[Lemma 2, Bent et al.~\cite{BentST85}]
    \label{lem:depth}
    Let $d$ be the depth of leaf $x$ in tree $T$ then $d < \lg(W_T/w_x) + 2$.
\end{lemma}

\begin{lemma}[Theorem 2, Bent et al.~\cite{BentST85}]
    \label{lem:join}
    Two biased trees $T$ and $S$ can be joined in amortized $| r(T) - r(S)|$ time.
\end{lemma}
Algorithm~\ref{algo:join} describes the algorithm of Bent et al. \cite{BentST85} that joins two trees. We describe it here because we will refer to details of this algorithm later. We refer the reader to the Bent et al.~\cite{BentST85} for the proofs of correctness
and complexity.

\begin{algorithm}[t]
   \DontPrintSemicolon
    \KwIn{Let $x$ and $y$ be the roots of the trees we are joining and assume without loss of generality
    that $r(x) \geq r(y)$.}
    \BlankLine
    \uIf(\tcc*[f]{Case 1}){$r(x) = r(y)$, or $r(x) > r(y)$ and $x$ is a leaf}{Create and return a new vertex with vertices $x$ and $y$ as its two children.}
    \ElseIf(\tcc*[f]{Case 2}){$r(x) > r(y)$ and $x$ is not a leaf}{ 
    Let $u$ be the right child of $x$.
    
    Remove $u$ as a child of $x$ and recursively join the trees with roots $u$ and $y$, producing
    a single tree, say with root $v$.
    
    \If(\tcc*[f]{Subcase 2a}){$r(v) \leq r(x) - 1$}{Attach $v$ as the right child of $x$ and return $x$.}
    
    \If(\tcc*[f]{Subcase 2b}){$r(v) = r(x)$}{In this case $v$ has exactly two children. Attach
        these as children of $x$ (to the right to the other children of $x$) and destroy $v$.
        Vertex $x$ thus gains a child.
        
        \lIf{$x$ has at most $3$ children}{return $x$.}
        \Else(\tcc*[h]{$x$ has $4$ children}) {
        Split $x$ into two vertices with $2$ children each, make them children of a new vertex $w$, and return $w$.        
        The two vertices resulting from the split has the same rank as $x$ while the rank of
        $w$ is one greater.}}
    }
\caption{Joining Biased Trees}
\label{algo:join}
\end{algorithm}



\begin{lemma}[Theorem 5, Bent et al.~\cite{BentST85}]
\label{lem:split}
The amortized time to split a tree $T$ at leaf $y$ is $r(T) - r(y)$.
\end{lemma}

\begin{lemma}[Theorem 6, Bent et al.~\cite{BentST85}]
\label{lem:split2}
The amortized time to split a tree $T$ at a key $i$ which is not in a leaf of $T$
is $O(\lg \frac{W_T}{\min(w_{i-}, w_{i+})})$ where $i^-$ and $i^+$ are the respective predecessor
and successor of $i$ in $T$.
\end{lemma}

\begin{lemma}[Theorem 7, Bent et al.~\cite{BentST85}]
\label{lem:reweigh}
The amortized time to change the weight of item $x$ in tree $T$ is $O(\lg \frac{\max(W_T, W'_T)}{\min(w_x, w'_x)})$
where $W_T, W_T', w_x, w'_x$ are the weights of the tree before and after the update and the weight
of $x$ before and after the update, respectively.
\end{lemma}

The lemmas above are proven by Bent et al.~\cite{BentST85} using the accounting method and require that
the trees satisfies the following \emph{credit invariant}: Every minor vertex $y$ with parent $x$ contains $r(x) - r(y) - 1$ credits.
Clearly, a singleton tree satisfies this invariant, and Bent et al. shows that any tree produced by combinations of the above
operations also does.

The \emph{c-rank} of a tree $T$ denoted $c(T)$ is the rank of $T$ plus the number of credits it has in its root,
i.e., $c(T) = j$ if the root of $T$ has $j - r(T)$ credits.

\noindent It follows from Lemma \ref{lem:join} that if two trees $S$ and $T$ have c-rank  $j > \max(r(S), r(T))$ then 
we can join them in $O(1)$ amortized time into a tree $U$ where $c(U) = j$.

\section{Data Structure}\label{sec:datastructure}
Our data structure maintains every set $G \in \mathcal{G}$
as a biased 2,3-tree \cite{BentST85}. 
We employ the weighting scheme, identical to the one used by
Iacono and \"{O}zkan~\cite{Iacono2010}, where the weight of a leaf $x \in G$ is $g^+(x) + g^-(x)$. 

Recall, that  $x^-$ and $x^+$ denote the predecessor and successor of $x \in G$
and $g^-(x) = 1$ if $x = \min(G)$ and $g^-(x) = x - x^-$ otherwise.
Similarly, $g^+(x) = 1$ if $x = \max(G)$ and $g^+(x) = x^+ - x$ otherwise.

\section{Biased Segment Merge}\label{sec:merge}
In this section we describe and analyze our merging algorithm. 
To obtain the desired complexity we will deviate from the sequential splitting
strategy of segment merge. Instead of finding the segments $A_i$ and $N_i$ and then joining them before finding the next segments, we will first find all the segments and then merge them. This way, we can avoid to use finger versions of the operations and this allows us to make an overall analysis of the cost of the $k$ merges. 

The biased segment merge algorithm has three main steps. First we split $A$ and $B$ into \emph{segments}
$\{ A_1, \ldots, A_k\}$  and $\{ B_1, \ldots, B_k\}$ where $A_i \subseteq A$ and $B_i \subseteq B$ and $\max(A_i) < \min(B_i)$ and $\max(B_i) < \min(A_{i+1})$ and construct a biased tree for each segment. 
This is done by first finding all the leaves correcponsding to the endpoints by doing a parallel search on the trees $A$ and $B$ and what we call the \emph{profile} of each segment (a profile of a segment consists of all the subtrees from the original tree that only contains leaves in the segment). Then the trees of each profile is joined into a biased segment tree. Finally we reweight the segment trees and join them into a single tree.

\begin{algorithm}[t]
\DontPrintSemicolon

{\bf Split} $A$ and $B$ into \emph{segments}
        $\{ A_1, \ldots, A_k\}$  and $\{ B_1, \ldots, B_k\}$ where $A_i \subseteq A$ and $B_i \subseteq B$ 
        and $\max(A_i) < \min(B_i)$ and $\max(B_i) < \min(A_{i+1})$ and construct a biased tree for each segment:
\MyBlock( ){
Find the set of profiles for all the segments:
\MyBlock{($A_1^l \leadsto A_1^r, \ldots, A_k^l \leadsto A_k^r$, $B_1^l \leadsto B_1^r, \ldots, B_k^l \leadsto B_k^r$) = {\sc FindProfiles}($A$,$B$)}
Construct the biased segment trees: 
\MyBlock{
\For{$i=1$ \KwTo $k$}{$A_i=$ {\sc ConstructSegmentTree}($A_i^l \leadsto A_i^r$)\;
$B_i=$ {\sc ConstructSegmentTree}($B_i^l \leadsto B_i^r$)
}
}
}

{\bf Reweight} the rightmost and leftmost leaves of $A_i$ and $B_i$: 
\MyBlock{ 
\For{$i=1$ \KwTo $k$}{
    $w(A_i^r) \leftarrow g^-(A_i^r) + a_i'$ 
    
    $w(A_i^l) \leftarrow g^+(A_i^r) + a_{i-1}''$
    
     $w(B_i^r) \leftarrow g^-(B_i^r) + b_i'$ \;
    
     $w(B_i^l) \leftarrow g^+(B_i^r) + b_{i-1}''$ \;
    }
}
{\bf Join} the segments $A_1, B_1, A_2, B_2, \ldots$ to produce a single biased tree $C$ by repeatedly joining the minimal rank tree with its minimal rank neighbor (solve ties arbitrarily).
\caption{Merge Biased Trees $A$ and $B$ into a single tree $C$}
\end{algorithm}

The full algorithm is described in Algorithm~2. 
In Section \ref{sub:splitting} we how to find the profiles and construct the segment trees.
In Section \ref{sec:complexity} we bound the combined amortized complexity of the biased segment merge.
\\\\
\noindent {\bf Note} For simplicity we will assume that $A\cap B = \emptyset$,  $\min(A) < \min(B)$, and $\max(A) < \max(B)$. We show in Section~\ref{sec:othercases} and \ref{sec:intersect} how to lift these assumptions.




\subsection{Splitting Into Segments}\label{sub:splitting}

Let $A_i^l$ and $A_i^r$ be the minimal and maximal element in segment $A_i$, respectively. Similarly, let $B_i^l$ and $B_i^r$ be respectively the minimal and maximal element in segment $B_i$.
Note that $A_1^l$ and $A_k^r$ and  $B_1^l$ and $B_k^r$ are the leftmost and rightmost leaves
of $A$ and $B$, respectively.  

Denote by $x \leadsto y$ the simple path from vertex $x$ to vertex $y$ in a tree $T$.
The \emph{profile} of a path $A_i^l \leadsto A_i^r$ are all the maximal subtrees of $A$ whose keys are in the range
 $(A_i^l;A_i^r)$ plus the leaves $A_i^l$ and $A_i^r$.
The root of each of these subtrees is a child of a vertex on the path $A_i^l \leadsto A_i^r$ except for the
subtrees $A_i^l$ and $A_i^r$. The \emph{profile} of a path $B_i^l \leadsto B_i^r$ is defined in the same way. We abuse notation and also call the \emph{profile} of a path $A_i^l \leadsto A_i^r$  the profile of the segment $A_i$.

\paragraph{Finding profiles} We find the profiles of the segments as 
described in Procedure~{\sc\ref{algo:split}}. 
Note that, rather than starting every search from the root, we continue from the leaf where the previous search ended (only the first search starts in the root). 



\begin{procedure}
\DontPrintSemicolon
    Find the leaves $A_1^l$ and $B_1^l$ by walking from the root to the leftmost leaf in $A$ and $B$, respectively.\;
    
    \For{$i = 1$ \KwTo $k-1$}{
        Find the leaf $A_i^r$ by searching for the predecessor of $B_i^l$ in $A$. The search is done by walking from $A_i^l$.\;
        
        From $A_i^r$ find the leaf $A_{i+1}^l$ which is the successor of $A_i^r$.\;
        
        Walking from $B_i^l$ find the key $B_i^r$ by searching for the predecessor of $A_{i+1}^l$ in $B$.\;
        
        From $B_i^r$ find the key $B_{i+1}^l$ which is the successor of $B_i^r$.\;
        }
        While walking in the trees $A$ and $B$ it is straightforward to identify the profiles of the segments: When walking up from a node $v$ add $v$'s right siblings to the profile. When walking down from a node $v$ add all $v$'s left siblings to the profile.\;
\caption{FindProfiles($A$,$B$)}\label{algo:split}
\end{procedure}


\paragraph*{Constructing the segment trees} We join the subtrees from each profile to produce a tree for each segment as described in Procedure~{\sc \ref{algo:segtree}}.
    
    

\begin{procedure}[h!]
\DontPrintSemicolon
\KwIn{The profile of $A_i$: $A_i^l \leadsto A_i^r =  v_1, v_2, \ldots, v_j, u_k, u_{k-1}, u_{k-2}, \ldots u_1$, where $v_j$ is the highest vertex on the path.}
\KwOut{Biased tree for the segment $A_i$.}
\BlankLine
Initially let $L_i$ be the singleton tree $v_1$.\;
\For{$m = 2$ \KwTo $j$:}
    { 
    \tcc{Case 1}
   \uIf{$v_{m-1}$ has two siblings $x$ and $y$ in the profile of segment $A_i$}
    {Remove $x$ and $y$ as children of $v_m$ and make them children of a new vertex $q$
    with rank $\max(r(x), r(y)) + 1$.\;
    Join the new tree rooted at $q$ to the tree $L_i$.}
    \tcc{Case 2}
    \ElseIf{$v_{m-1}$ has a single sibling $q$ in the profile of segment $A_i$}{
    Remove $q$ as a child of $v_m$\;
    Join the tree rooted at $q$ to $L_i$.}
    
}
    
    

     Produce $R_i$ symmetrically by considering 
    the vertices $u_1, \ldots, u_k$.\;
    Join $L_i$ and $R_i$ to obtain $A_i$.\;
    \KwRet{$A_i$}
    
\caption{ConstructSegmentTree($A_i^l \leadsto A_i^r $)}\label{algo:segtree}
\end{procedure}

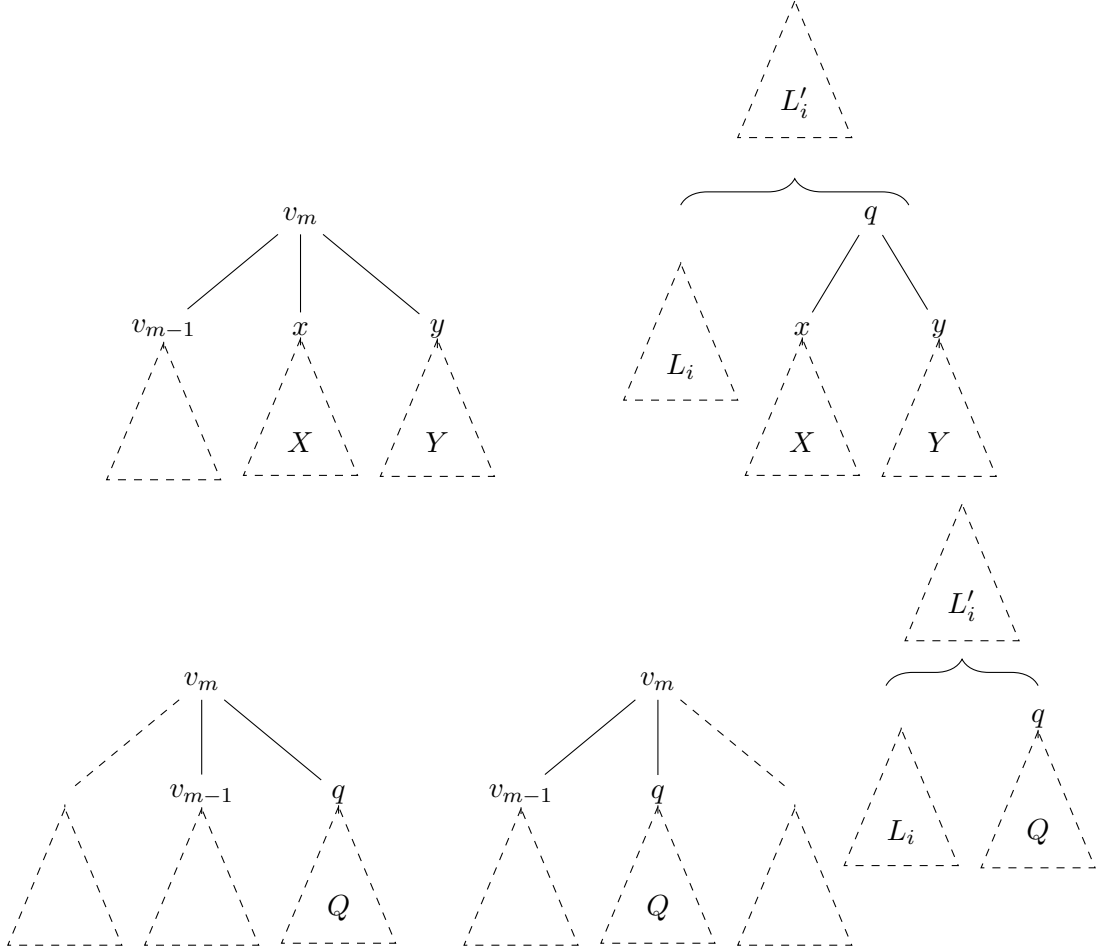
\begin{figure}[htb]
\begin{center}
    \begin{tikzpicture}[sibling distance=1.8cm]
    \node {$v_m$}
            child{ node {$v_{m-1}$}
                child { node[triangle] {} }
            }
            child{ node {$x$}
                child { node[triangle] {$X$} }
            }
            child{ node {$y$}
                child { node[triangle] {$Y$} }
            };
            
    \node at (5.0, -0.5) { $$}
        child { node[triangle] {$L_i$} };  
        
    \node at (7.5, 0) { $q$}
        child { node { $x$ } 
            child { node[triangle] {$X$} }
        }
        child { node { $y$ } 
            child { node[triangle] {$Y$} }
        };
        
    \node at (6.5, 3.0) { $$}
        child { node[triangle] {$L'_i$} };
        
    \draw [decorate,decoration={brace,amplitude=10pt,raise=4pt},yshift=0pt]
    (5,0) -- (8,0) node [black,midway,xshift=0.8cm]  {};
    \end{tikzpicture}

    \begin{tikzpicture}[sibling distance=1.8cm]
        \node {$v_m$}
                child[dash]{ node {$$}
                    child { node[triangle] {} }
                }
                child{ node {$v_{m-1}$}
                    child { node[triangle] {} }
                }
                child{ node {$q$}
                    child { node[triangle] {$Q$} }
                };
        \node at (6, 0) {$v_m$}
                child{ node {$v_{m-1}$}
                    child { node[triangle] {} }
                }
                child{ node {$q$}
                    child { node[triangle] { $Q$} }
                }
                child[dash]{ node {$$}
                    child { node[triangle] {$$} }
                };
                
    \node at (10.0, 2.5) { $$}
        child { node[triangle] {$L'_i$} };
        
    \node at (9.2, -0.5) { $$}
        child { node[triangle] {$L_i$} };

    \node at (11, -0.5) { $q$}
        child { node[triangle] {$Q$} };
        
    \draw [decorate,decoration={brace,amplitude=10pt,raise=4pt},yshift=0pt]
    (9,-0.2) -- (11,-0.2) node [black,midway,xshift=0.8cm] {};
            
    \end{tikzpicture}
\caption{Case 1 and 2 from Procedure~{\sc\ref{algo:segtree}}.}   
\end{center}
\end{figure}


To prove that the algorithm correctly produce a biased tree for each segment $A_i$ we need to 
verify that the credit and bias invariant are true. It follows from Bent~et~al.\ that the join and split operations maintain both invariants. Thus we only need to prove that the new tree rooted in  $q$ in case 1 satisfy both the invariants. In case 1 at least one of $x$ and $y$ are major (since the original tree satisfies the bias invariant) and $r(q)=r(v_m)$. If either $x$ or $y$ is minor then the other one is a major leaf (since the original tree satisfies the bias invariant) and thus the tree rooted at $q$ is biased.  If $y$ is minor then $y$ has $r(v_m)-r(y) + 1$ credits due to the credit invariant in the original tree. Since $r(v_m)=r(q)$ this implies that the credit variant is satisfied in the tree rooted at $q$. The argument when $x$ is minor is similar. 





\subsection{Biased Merge Complexity}\label{sec:complexity}

The cost of performing the biased segment merge
is the sum of the cost of splitting $A$ and $B$ into segments,
reweighting the segments, and joining them into a single tree. 

\subsubsection{Cost of constructing segment trees} 
Consider the vertices of $A$ that are visited in the search for the segment endpoints. 
This search visits vertices on the path $A_i^l \leadsto A_i^r$ for $i = 1,\ldots, k$ along with any ancestor of such
vertex at most $O(1)$ times.   It follows from the definition of a profile that the number of trees in the profile is linear in the length of the path.  Having found the path $A_i^l \leadsto A_i^r$ we can find the roots of the trees in the profile in linear
time in the length of the path.
After having removed all subtrees that are in the profiles from $A$, the remaining vertices are either branching, or
were a parent of a subtree in the profile. The number of vertices visited during this traversal therefore is no  more than the total number of substrees in the profile which is no more than the total number of joins.

 Let $W_{x \leadsto y}$ be the combined weight of the subtrees in the profile of $x \leadsto y$.
 Observe that the keys in the profile of the path $A_i^l \leadsto A_i^r$ are exactly the keys of $A_i$
 and thus $W_{A_i^l \leadsto A_i^r} = W_{A_i}$.

Before analyzing the cost of the joins we need some definitions. The following definition is similar to the one in Iacono and {\"O}zkan~\cite{Iacono2010}.

\begin{figure}
	\begin{center}
\begin{tikzpicture}[x=0.48cm,y=2cm,decoration=brace]
        \draw[-] (-5,0)--(24,0);
        
        \foreach \x in {-5,...,-2}
        \fill (\x,0) circle (2pt);
        \fill (-3.5,0) circle (2pt);

        \foreach \x in {5,...,9}
        \fill (\x,0) circle (2pt);
        \fill (6.5,0) circle (2pt);
        \fill (9.5,0) circle (2pt);
        
        \foreach \x in {17,...,19}
        \fill (\x,0) circle (2pt);
        \fill (17.5,0) circle (2pt);
          
        \draw[decorate, yshift=2ex]  (-5,0) -- node[above=0.4ex] {$A_{i-1}$}  (-2,0);
        
        \draw[decorate, yshift=2ex]  (5,0) -- node[above=0.4ex] {$A_{i}$}  (9.5,0);
        
        \draw[decorate, yshift=2ex]  (17,0) -- node[above=0.4ex] {$A_{i+1}$}  (19,0);
        
        \draw[|-|,yshift=-2ex] (-1.8,0)--(4.8,0);
        \node[below,yshift=-2ex] at (1.5,0) {$a_{i-1}$};
        
        \draw[|-|,yshift=-2ex] (9.7,0)--(16.8,0);
        \node[below,yshift=-2ex] at (13,0) {$a_{i}$};
        
        \draw[|-,yshift=-2ex] (19.1,0)--(24,0);
        \node[below,yshift=-2ex] at (21,0) {$a_{i+1}$};
        
        \node[left] at (-5,0) {$A:\;\ldots$};
        \node[right] at (24,0) {$\ldots$};
        
        \draw[-] (-5,-1)--(24,-1);
        
        \foreach \x in {0.5,...,2.5}
        \fill (\x,-1) circle (2pt);
        \fill (2,-1) circle (2pt);

        \foreach \x in {12,...,13}
        \fill (\x,-1) circle (2pt);
        \fill (12.5,-1) circle (2pt);
        
        \foreach \x in {21,...,23}
        \fill (\x,-1) circle (2pt);
        \fill (23.5,-1) circle (2pt);
          
        \draw[decorate, yshift=2ex]  (0.5,-1) -- node[above=0.4ex] {$B_{i-1}$}  (2.5,-1);
        
        \draw[decorate, yshift=2ex]  (12,-1) -- node[above=0.4ex] {$B_{i}$}  (13,-1);
        
        \draw[decorate, yshift=2ex]  (21,-1) -- node[above=0.4ex] {$B_{i+1}$}  (23.5,-1);
        
        \draw[-|,yshift=-2ex] (-5,-1)--(0.3,-1);
        \node[below,yshift=-2ex] at (-3,-1) {$b_{i-2}$};
        
        \draw[|-|,yshift=-2ex] (2.7,-1)--(11.8,-1);
        \node[below,yshift=-2ex] at (7.25,-1) {$b_{i-1}$};
        
        \draw[|-|,yshift=-2ex] (13.2,-1)--(20.8,-1);
        \node[below,yshift=-2ex] at (17,-1) {$b_{i}$};
        
        \node[left] at (-5,-1) {$B: \;\ldots$};
        \node[right] at (24,-1) {$\ldots$};
        

         \draw[-] (-5,-2)--(24,-2);
        
        \foreach \x in {-5,...,-2}
        \fill (\x,-2) circle (2pt);
        \fill (-3.5,-2) circle (2pt);

        \foreach \x in {5,...,9}
        \fill (\x,-2) circle (2pt);
        \fill (6.5,-2) circle (2pt);
        \fill (9.5,-2) circle (2pt);
        
        \foreach \x in {17,...,19}
        \fill (\x,-2) circle (2pt);
        \fill (17.5,-2) circle (2pt);  
        
        \foreach \x in {0.5,...,2.5}
        \fill (\x,-2) circle (2pt);
        \fill (2,-2) circle (2pt);

        \foreach \x in {12,...,13}
        \fill (\x,-2) circle (2pt);
        \fill (12.5,-2) circle (2pt);
        
        \foreach \x in {21,...,23}
        \fill (\x,-2) circle (2pt);
        \fill (23.5,-2) circle (2pt);
        
        
        \draw[|-|,yshift=-2ex] (-1.9,-2)--(0.4,-2);
        \node[below,yshift=-2ex] at (-0.75,-2) {\footnotesize{$a'_{i-1}=b''_{i-2}$}};
        
        \draw[|-|,yshift=-2ex] (2.6,-2)--(4.9,-2);
        \node[below,,yshift=-2ex] at (3.75,-2) {\footnotesize{$a''_{i-1}=b'_{i-1}$}};
        
        \draw[|-|,yshift=-2ex] (9.6,-2)--(11.9,-2);
        \node[below,,yshift=-2ex] at (10.75,-2) {\footnotesize{$a'_{i}=b''_{i-1}$}};
        
        \draw[|-|,yshift=-2ex] (13.1,-2)--(16.9,-2);
        \node[below,yshift=-2ex] at (15,-2) {\footnotesize{$a''_{i}=b'_{i}$}};
        \draw[|-|,yshift=-2ex] (19.1,-2)--(20.9,-2);
        \node[below,yshift=-2ex] at (20,-2) {\footnotesize{$a'_{i+1}=b''_{i}$}};
        
        \node[left] at (-5,-2) {$C: \;\ldots$};
        \node[right] at (24,-2) {$\ldots$};

\end{tikzpicture}
\end{center}
	\caption{Illustration of segments and the surrounding gaps. These gaps are used for both weighting and as potential when analyzing complexity.}
	\label{fig:merge}
\end{figure}
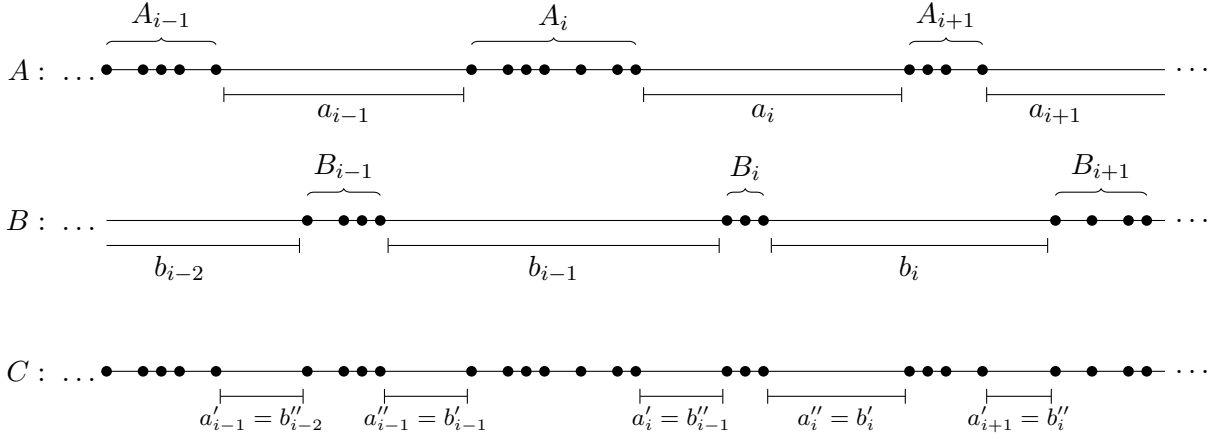

\paragraph{Gaps} For $i = 1, \ldots, k-1$ let $a_i = A_{i+1}^l - A_i^r$ and $ b_i = B_{i+1}^l - B_i^r$. That is, $a_i$ and $b_i$ are the length of the gap between the $i^{th}$ and the $(i+1)^{th}$ segment in $A$ and $B$, respectively (see Figure~\ref{fig:merge}.).  Define $a_0 = b_k = 1$. We will abuse notation and let $a_i$, resp.\ $b_i$, denote both the gap and the length of the gap. 

After the merge the gap $a_i$ is intersected by the segment $B_i$ splitting the gap $a_i$ into two smaller pieces $a'_i$ and $a''_i$. The gap $b_i$ is split similarly into two gaps $b'_i$ and $b''_i$ by $A_{i+1}$. Thus, $a_i'' = b_i' = A_{i+1}^l - B_i^r$ and $a_i' = b_{i-1}'' = B_{i}^l - A_i^r$. We define $a''_0 = b'_k = 1$.

%
We need the following property: 

\begin{property}\label{prop:intervals} The weight of the segment trees can be bounded as follows:
\begin{itemize}
    \item $W_{A_i} \leq a_{i-1} + a_{i} + 2b_{i-1}$ for $i = 2, \ldots, k$ and
    \item $W_{B_i} \leq b_{i-1} + b_{i} + 2a_i$ for $i = 1, \ldots, k-1$.
\end{itemize}
\end{property}
The weight of the segment tree $A_i$ before the reweighting is two times the sum of the gaps in $A_i$ + $a_{i-1}$ + $a_i$. The sum of the gaps in $A_i$ is at most $b_i$ and the property follows. The argument for $B_i$ is similar. 

\paragraph{Cost of the joins} We now consider the amortized complexity of the joins performed on $L_i$ in a single iteration.
The case for $R_i$ is analogous, and the analysis is similar to the analysis of the split operation of Bent et al.~\cite{BentST85}.
Observe that $v_{m-1}$ is an ancestor to all the trees we have joined into $L_i$ before this iteration. 
Let $L_i'$ be the tree $L_i$ after this iteration, and assume that $r(L_i) \leq r(v_{m-1})$ and that the root of $L_i$ has c-rank $r(v_{m-1}) + 1 $.
We will prove that $r(v_m) - r(v_{m-1}) + 3$ new credits suffice to both pay for the iteration and leave enough credits in the root of 
$L'_i$ to make $c(L'_i) = r(v_m) + 1$ and that $r(L_i') \leq r(v_m)$ making our assumptions true for the next iteration.

In case 1 at least one of $x$ and $y$ are major (because of bias) and thus $r(q) = r(v_m)$.
If we place $r(v_m) - r(v_{m-1})$ new credits in the root of $L_i$ and $1$ new credit on $q$ we have $c(L_i) = r(v_m) + 1 = c(q)$.
In case 2, if $q$ is minor then it has $r(v_m) - r(q) - 1$ credits by the credit invariant, otherwise it is major and thus $r(q) = r(v_m) - 1$.
In any case $c(q) = r(v_m) - 1$. If we place $2$ new credits on $q$ and $r(v_m) - r(v_{m-1})$ new credits in the root of $L_i$
then again $c(L_i) = r(v_m) + 1 = c(q)$.
In case 3, we place $r(v_m) - r(v_{m-1})$ credits in the root of $L_i$ and thus $c(L'_i) = r(v_m)$.

In all three cases we spend at most $r(v_m) - r(v_{m-1}) + 2$ new credits, in cases 1 and 2,
the join of $L_i$ and $q$ takes $O(1)$ amortized time.
Also, there is at least $1$ unused new credit which pays for the $O(1)$ work of every iteration
and  after the iteration $c(L'_i) = r(v_m) + 1$.

We now prove that our assumption $r(L_i) \leq r(v_{m-1})$ holds.
Initially, $L_i = v_1$ and thus $r(L_i) = r(v_1)$. We will prove that if $r(L_i) \leq r(v_{m-1})$
then $r(L'_i) = r(v_m)$ which makes our assumption true for the next iteration. 
In case 1, either $x$ or $y$ is major by local bias and thus $r(q) = r(v_m)$.
This means that when joining $q$ and $L_i$ we end up in case $2$ of Algorithm~\ref{algo:join}.
By inspecting cases $2a$ and $2b$ of Algorithm \ref{algo:join} and using the fact that
$q$ has two children, it is clear that $q$ will be the root of $L'_i$ and
thus $r(L'_i) = r(v_m)$.
In case 2 we are joining the tree rooted at $q$ and $L_i$
and because $r(q) \leq r(v_m)$ and $r(L_i) \leq r(v_{m-1}) < r(v_m)$
it follows from Algorithm~\ref{algo:join} that $r(L'_i) \leq r(v_m)$.
In case 3, $L'_i = L_i$ and since $r(L_i) \leq r(v_{m-1})$
it follows that $r(L'_i) \leq r(v_{m})$.

Now let's look at the overall cost.
The amortized cost of a single step of the algorithm is proportional to the rank difference between
consecutive vertices on the path and thus the sum of these costs telescope and 
the total cost becomes $r(v_j) - r(v_1)$, where $v_j$ is the maximal rank tree
merged into $L_i$ and $v_1 = A_i^l$. By the definition of rank and Lemma~\ref{lem:max-rank} we have  $r(v_j) = O(\lg w_{v_j}) = O(\lg W_{A_i})$  and 
we also have $\lg w_{A_i^l}\geq \lg a_{i-1}$. 
Therefore, the cost of of constructing $L_i$ is $O(\frac{\lg W_{A_i}}{ \lg a_{i-1}})$.
By a similar argument the cost of of constructing $R_i$ is $O(\frac{\lg W_{A_i}}{ \lg a_{i}})$. 
The amortized complexity of constructing $L_i$ and $R_i$ is thus $O(\lg \frac{W_{A_i}}{\min(a_{i-1}, a_i)})$. 
Since $a_{i-1} \leq r(L_i) \leq w(A_i))$ and $a_{i} \leq r(R_i) \leq w(A_i))$ the complexity of joining $L_i$ and $R_i$ is also bounded by this, and thus
it also bounds the complexity of constructing the segment tree $A_i$.
Summing over all the segments of $A$ this also bounds the number of vertices visited during the traversal.
In summary we have proven the following lemma:
\begin{lemma}
The complexity of constructing the trees $A_i, B_i$ for $i = 1, \ldots, k$ is
$$
    O\left(\sum_{i=1}^{k} \left(\lg \frac{W_{A_i}}{\min(a_{i-1}, a_i)} + \lg \frac{W_{B_i}}{\min(b_{i-1}, b_i)}\right)\right)\;.
$$
\end{lemma}

\subsubsection{Cost of reweighting}\label{sec:complexity-reweight}
By Lemma~\ref{lem:reweigh} the total complexity of reweighting $A_i$ is 
$O(\lg \frac{W_{A_i}}{a_i'} +  \lg \frac{W_{A_i}}{a_{i-1}''} )$, 
and total reweighting cost for all the segments is then 

$$
    O\left(\sum_{i=1}^k \left(\lg \frac{W_{A_i}}{a_i'} +  \lg \frac{W_{A_i}}{a_{i-1}''} + \lg\frac{W_{B_i}}{b_i'} +  \lg \frac{W_{B_i}}{b_{i-1}''}\right)\right) = 
$$
$$
    O\left(\sum_{i=1}^k \left(\lg \frac{W_{A_i}}{\min(a_i', a_{i-1}'')} + \lg\frac{W_{B_i}}{\min(b_i', b_{i-1}'')}\right)\right)\;.
$$

\subsubsection{Cost of joining the segment trees}
Observe that there are exactly $2k-1$ joins.
We bound the complexity of joining the segments $A_1, B_1, A_2, B_2, \ldots, A_k, B_k$ by considering
the merge tree resulting from the sequence of joins in the algorithm. 
This is a binary tree with $2k-1$ vertices where the leaves in 
left-to-right order represent the segments in order and an internal vertex represents the tree given by joining
its two children. We assign to each internal vertex the cost of joining its two children which by Lemma~\ref{lem:join} is the difference in the rank of the two children.
Thus the complexity of the join algorithm is no more than the sum of the costs assigned to the vertices in the merge tree.

We need the following fact for the analysis, which is easily verified by considering the different cases of Algorithm \ref{algo:join}.
\begin{property}\label{prop:rank}
Let $T$ be the tree obtained by joining two trees $P$ and $S$.
Then $T$ has rank $r(T) \in \{\max(r(P), r(S)), 1 + \max(r(P), r(S))\}$.
\end{property}

First consider an internal vertex in the merge tree where 
the difference in rank between its children is at most $2$.
The sum of the cost assigned to such vertices is at most $2(2k-1) = O(k)$.
Now consider an internal vertex $v$ assigned a cost $c_v > 2$.
Let $a$ and $b$ be the children of $c_v$ 
and assume without loss of generality that $c_v = r(b) - r(a)$ and thus $r(b) > r(a) + 2$.
We prove that either $b$ is a leaf or the left child of $b$ is a leaf
with rank at least $r(b) - 1$.  

\begin{figure}[h!]
\centering
\begin{tikzpicture}[nodes={draw, circle, minimum size=1cm}, ->]
 
 \node{ $v$ }
    child { node {$a$} }
    child { node {$b$} 
        child { node {$c$}
            child { node {$e$} }
            child { node {$f$} }
        }
        child { node {$d$} }  
    };
\end{tikzpicture}
	\caption{Part of the subtree rooted at $v$.}
	\label{fig:tree}
\end{figure}
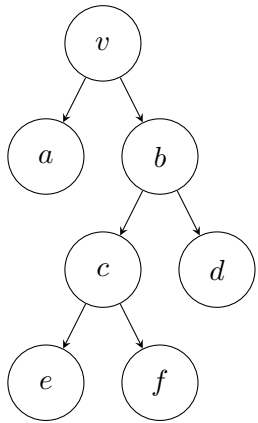

If $b$ is a leaf, then we are done, so let $c$ and $d$ be its children as depicted in Figure \ref{fig:tree}.
We must have $r(d) \leq r(a)$, otherwise, $a$ and $c$ would have been joined before $c$ and $d$.
As $r(b) > r(a) + 2$ and $r(d) \leq r(a)$ we must have $r(c) > r(a) + 1$ and $r(c) \geq r(b) - 1$ by Property~\ref{prop:rank}.
If $c$ is a leaf then we are done, so assume that $c$ has children $e$ and $f$ as shown in Figure~\ref{fig:tree}.
Because $r(c) > r(a) + 1$  either $r(e) > r(a)$ in which case $f$ and $d$ would have been joined before $e$ and $f$,
or $r(f) > r(a)$ in which case $a$ and $e$ would have been joined before $e$ and $f$.
In either case we have reached a contradiction,
so we have proved that either $b$ is a leaf or $c$ is a leaf of rank at least $r(b) - 1$.
The argument for $r(a) > r(b) + 2$ is symmetric.

Observe that $r(a)$ is at least as great as the rank of the rightmost leaf in the left subtree
of $v$ and that the leftmost leaf in the right subtree of $v$ has rank $r(b)$ or rank $r(b)-1$.
These leaves corresponds to consecutive segments $A_i, B_i$ or $B_i, A_{i+1}$ for some $i$,
and it follows that the cost $c_v$ is no greater than the cost of joining these two segments.
If we charge the cost to these two segments, it is easy to see that no other vertex $v'$ with
cost $v'_c$ will be charged to the same two same segments, thus the total cost
is at most 
$$
    O\left(\sum_{i=1}^k \left|r(A_i) - r(B_i)\right| + \sum_{i=1}^{k-1} \left|r(B_i) - r(A_{i+1})\right|\right)\;.
$$

Recall that $W_{A_{i+1}} \geq a_i \geq a''_i = b'_i$ and $W_{B_i} \geq  b_{i-1} \geq b''_{i-1} = a'_i$ which implies that

\begin{align}
    &O\left(\sum_{i=1}^k |r(A_i) - r(B_i)| + \sum_{i=1}^{k-1} |r(B_i) - r(A_{i+1})|\right)\\
    =\;&O\left(\sum_{i=1}^k (\lg W_{A_i} - \lg a'_i) + \sum_{i=1}^{k-1} (\lg W_{B_i} - \lg b'_i)\right)\\
    =\;&O\left(\sum_{i=1}^k \left(\lg \frac{W_{A_i}}{a'_i} + \lg \frac{W_{B_i}}{b'_i}\right)\right)\;.
\end{align}

\subsubsection{Amortized complexity of biased segment merge}
The total cost of performing the biased segment merge
is the sum of the cost of splitting $A$ and $B$ into segments,
reweighting the segments, and joining them into a single tree as described in the previous sections.
Thus the total complexity is dominated by the reweighting of the segments endpoints which takes time
\begin{align}
& O\left(\sum_{i=1}^k \left(\lg \frac{W_{A_i}}{\min(a_i', a_{i-1}'')} + \lg \frac{W_{B_i}}{\min(b_i', b_{i-1}'')}\right)\right)\\
=\;& O\left(\lg W_{A_1} + \lg W_{B_1} + \lg W_{A_k} + \lg W_{B_k} + \sum_{i=2}^{k-1}\left( \lg \frac{W_{A_i}}{\min(a_i', a_{i-1}'')} +  \lg \frac{W_{B_i}}{\min(b_i', b_{i-1}'')}\right)\right)\\  \label{l9}
=\;&O\left(\lg W_A + \lg W_B + \sum_{i=2}^{k-1} \left(\lg \frac{2b_{i-1} + a_i + a_{i-1}}{\min(a_i', a_{i-1}'')} + \lg \frac{2a_i + b_i + b_{i-1}}{\min(b_i', b_{i-1}'')}\right)\right)\\
 \label{l10}
=\;&O\left(\lg W_A + \lg W_B  + \sum_{i=2}^{k-1} \left(\lg \frac{\max(b_{i-1},a_i,a_{i-1})}{\min(a_i', a_{i-1}'')} + \lg \frac{\max(a_{i}, b_{i}, b_{i-1})}{\min(b_{i}', b_{i-1}'')}\right)\right)\\\label{l11}
=\;&O\left(\lg W_A + \lg W_B  + \sum_{i=2}^{k-1} \left(\lg \frac{\max(a_i, a_{i-1}, b_i, b_{i-1})}{\min(a_i', a_{i-1}'', b_i',b''_{i-1})}\right)\right)
\end{align}
where (\ref{l9}) follows from 
Property~\ref{prop:intervals} and (\ref{l10}) follows from the relations between the gap sizes. 

\paragraph{Potential function} We now analyze the amortized cost of this operation using the following potential function:
Define the potential of a set $G$ as $$\phi(G) = \sum_{x \in G} (\lg g^+(x) + \lg g^-(x))$$ 
like before and
let the potential function of the mergeable dictionary be 
$$
    \Phi(\mathcal{G}) = \sum_{G \in \mathcal{G}} \phi(G)\;.
$$

The potential of an empty data structure is $0$ and clearly the potential remains non-negative.
When performing a merge between two sets $A, B \in \mathcal{G}$, only
the potential of elements in $A$ and $B$ are affected. Furthermore, only the minimum and maximum
element in each of the segments $A_i, B_i$ are affected.
The size of the right gap of $A_i^r$ changes from $a_i$ to $a_i'$
and the size of the left gap of $A_i^l$ changes from $a_{i-1}$ to $a_{i-1}''$. All gap sizes decrease except the left gap of $B_1^l$ and the right gap of $A_k^r$.
The total potential \emph{decrease} is therefore at least
\begin{align}
     &\sum_{i=2}^{k-1} \left(\lg a_i - \lg a_i' + \lg a_{i-1} - \lg a_{i-1}'' + \lg b_i - \lg b_i' +  \lg b_{i-1} - \lg b_{i-1}''\right)\\    \label{l2}
   =\;& \Omega\left(\sum_{i=2}^{k-1} \lg \frac{\max(a_i, a_{i-1}, b_i, b_{i-1})}{\min(a_i', a_{i-1}'', b_i',b''_{i-1})}\right)\\\label{l1}
\end{align}
where  
(\ref{l2}) follows by the relations between the gap sizes. The potential increase is at most $$(\lg b''_0 - \lg b_0) + (\lg a'_k -\lg a_k)  \leq \lg a_0 - 0 + \lg b_k - 0 =\lg a_0 +\lg b_k \leq \lg W_A + \lg W_B.$$

It follows that the cost of the merge plus the change in potential
is no more than $O(\lg W_A + \lg W_B) = O(\lg W_C)$.
Therefore, the amortized cost of the merge operation is $O(\lg U_C)$
where $U_C$ is $\max(B) - \min(A) = \max(C) - \min(C)$.


\subsection{Lifting Assumptions}\label{sec:othercases} Here we assumed that $\min(A) < \min(B)$. It this is not the case, swap the arguments of the merge.
We also assumed that $\max(A) < \max(B)$. If this is not the case, we start by
executing $B', B'' \leftarrow \esplit{B}{\max(A)}$, we then perform $C' \leftarrow \emerge{A}{B'}$,
change the weight of the elements $\max(C')$ and $\min(B'')$ according to our weighting scheme 
and finally because $\max(C') = \max(A) < \min(B'')$ we can produce $C$ by joining $C'$ and $B''$.
It follows from Lemmas \ref{lem:max-rank}, \ref{lem:join}, \ref{lem:split2} and \ref{lem:reweigh}
that the total cost excluding the \emerge{A}{B'} operation is $O(\lg U_C)$. 

We show how to handle the case where $A\cap B\neq \emptyset$ in Section~\ref{sec:intersect}.

\section{Split, MakeSet, Search, and Shift}\label{sec:other-operations}  

\subsection{Supporting Shift} 

In order to support the \dshift{} operation,  we extend the biased trees by storing an integral offset with every vertex $x$
denoted $s(x)$ indicating that the subtree rooted at $x$ is shifted by $s(x)$.
We then maintain the \emph{shift invariant}: The key represented by a leaf $y$ in a tree $T$
is the sum of the offsets on the path from the root of $T$ to $y$.

Let $T_G$ be the biased tree representing the ordered set $G$.
The set $G$ is shifted by incrementing $s(x)$ by $j$ where $x$ is the root of $T_G$.

The operation \emakeset{j} creates a tree with a single vertex $x$ and sets $s(x):= j$.
During navigation, whenever visiting an internal vertex $x$ with $s(x) \neq 0$
we increment the values $x_l, x_r$ and the shift of each of its children by $s(x)$ and afterwards set $s(x): = 0$.
This gives a constant time overhead at each step and maintains the shift invariant.

Because all operations on a biased tree start at the root of $T_G$ it is then guaranteed
that if we navigate to a vertex $x$, then the shift invariant
is also true for the subtree rooted at $x$.
Furthermore, since the only way to manipulate a biased tree is by manipulating
its pointers, it follows that the shift invariant remains true under any operation.

\subsection{Analysis}
The total weight of the tree representing a dictionary $G$ is $O(\lg U_G)$ where $U_G = \max(G) - \min(G)$.

\paragraph{Search} The potential is not affected by the \dsearch{} operation which therefore by
Lemma \ref{lem:depth} takes $O(\lg U_G) = O(\lg U_G)$ worst-case and amortized time.

\paragraph{Split} The $A, B \leftarrow \esplit{G}{x}$ operation takes $O(\lg U_G)$ worst-case time by Lemma \ref{lem:split2} and Lemma \ref{lem:max-rank}. The size of the left gap of $A^r$ and the right gap of $B^l$ decrease and no
other gaps are affected, thus the amortized time of the operation is no worse than the worst-case time.

\paragraph{MakeSet} The \dmakeset{} operation creates a new set $G$ with a singleton element.
Creating the biased tree with a singleton vertex takes $O(1)$ worst-case time and requires no new credits.
As both the left and right gap of the element is $1$, the potential change is $O(1)$.
The amortized time of the operation is therefore no more than the worst-case time.

\paragraph{Shift}
The potential is not affected by the \dshift{} operation. Thus the worst case and amortized time of \dsplit{} is $O(1)$.



\section{Intersecting Segments}\label{sec:intersect}

We now consider how to handle the case where the sets $A$ and $B$ are not disjoint.
We will produce the set $C = A \cup B$ having only unique elements and can therefore inductively assume that $A$ and $B$ each consist of unique elements.

Observe that our technique for identifying and splitting segments does not
have any dependency on the keys of $A$ and $B$ being disjoint.
However, we can now only guarantee the inequalities 
 $\max(A_i) \leq \min(B_i)$ for $i = 1,\ldots, k$ and $\max(B_i) \leq \min(A_{i+1})$ for $i = 1, \ldots, k-1$.
 
We add a pruning step in between the splitting and reweighting. This is described in Procedure~{\sc\ref{algo:prune}}.
 
Keeping track of the multiplicity of every key can easily be obtained by storing a counter in every leaf. This only implies a constant overhead in the running time. 
 
\begin{procedure}[t]
    \DontPrintSemicolon
    \For{$i=1$ \KwTo $k$}{
        \If{$\max(A_i)=\min(B_i)$}{.
            \uIf{$A_i$ is a singleton tree and $B_i$ is not a singleton tree}{Delete $A_i$}
            \uElseIf{$B_i$ is a singleton tree}{Delete $B_i$}
            \Else {
                Split $\max(A_i)$ from $A_i$ \; \label{cl:s1}
	            Delete the new singleton tree containing $\max(A_i)$.
                }
            }
        \If{$\max(B_i)=\min(A_{i+1})$}{
            \uIf{$B_i$ is a singleton tree and $A_{i+1}$ is not a singleton tree}{Delete $B_i$}
            \uElseIf{$A_{i+1}$ is a singleton tree}{Delete $A_{i+1}$}
            \Else {
                Split $\max(B_i)$ from $B_i$ \; \label{cl:s2} 
	            Delete the new singleton tree containing $\max(B_i)$. \label{cl:d2} 
                }
            }
    }
    \caption{PruneTrees($A_1,\ldots,A_k$,$B_1,\ldots,B_k$)}\label{algo:prune}
\end{procedure}


 
\subsection{Analysis} 
We now analyze the amortized cost of the new merge operation. We first bound the cost of the pruning procedure.

\paragraph{Cost of pruning}
The cost of running through all the segment trees and checking if they overlap is bounded by the cost of constructing the segment trees.
The same is true for the deletion of the singleton trees. Note that the cost of deleting a singleton tree $A_i$ is $O(1)$ and causes the potential to decrease by $\lg W_{A_i}$.

It remains to account for the cost of the splits in line \ref{cl:s1} and \ref{cl:s2}. By Lemma~\ref{lem:split} the cost of splitting $\max(A_i)$ from $A_i$ is 
$$ O\left(\lg\frac{W_{A_i}}{w(\max(A_i))} \right) =  O\left(\lg\frac{W_{A_i}}{a_i} \right) = O\left(\lg\frac{W_{A_i}}{\min(a_i',a_{i-1}'')} \right) \;,
$$
since $w(\max(A_i) \geq a_i \geq \min(a'_i,a''_{i-1})$.
By similar arguments the cost of splitting $\max(B_i)$ from $B_i$ is $O\left(\lg\frac{W_{B_i}}{\min(b_i',b_{i-1}'')} \right)$.

We note that the new maximum elements in the trees where we deleted the maximum now already have the correct weight, and we will therefore not reweight these during the reweighting step. 
The cost of the reweighting for all other elements is the same as in the previous analysis (see Section~\ref{sec:complexity-reweight}). That is, the cost of reweighting $A_i^l$ is $O(\lg(W_{A_i}/a'_i))$ and the cost of reweighting $A_i^r$ is $O(\lg(W_{A_i}/a''_{i-1}))$, etc. 
 
The total cost of pruning and reweighting is therefore
$$
O\left(\sum_{i=1}^k \left(\lg \frac{W_{A_i}}{\min(a_i', a_{i-1}'')} + \lg\frac{W_{B_i}}{\min(b_i', b_{i-1}'')}\right)\right)\;.
$$




\paragraph{Amortized cost of merge}

We will argue that the potential still decreases with at least 
$$
\Omega\left(\sum_{i=2}^{k-1} \lg \frac{\max(a_i, a_{i-1}, b_i, b_{i-1})}{\min(a_i', a_{i-1}'', b_i',b''_{i-1})}\right)\;.
$$
Consider segment $A_i$ for some $2\leq i\leq k-2$. 
There are 5 cases: 
\begin{itemize}
    \item If $A_i$ is deleted then we get a potential loss of $\lg a_i + \lg a_{i-1}$.
    \item If the weight of $A_i^l$ decreases then we the potential decreases by $\lg a_{i-1} - \lg a''_{i-1}$. 
    \item If the weight of $A_i^r$ decreases then we the potential decreases by $\lg a_{i} - \lg a'_{i}$. 
    \item If the weight of $A_i^l$ does not decrease (and $A_i$ is not deleted) then $\min(B_{i-1}) = \max(A_{i-1})$ and either $B_{i-1}$  is a singleton set and $ \min(B_{i-1}) =  \max(B_{i-1})$ or $B_{i-1}$ consists of exactly two elements and $\max (B_{i-1}) = \min(A_i)$. If $B_i$ is a singleton set then $b_{i-1} \geq a_{i-1}$ and $B_{i-1}$ is deleted giving a potential decrease of at least $\lg b_{i-1} \geq \lg a_{i-1}$. Otherwise, $B_{i-1}^r$ is deleted (in line \ref{cl:d2}) giving potential decrease of at least $\lg a_{i-1}$. 
    \item If $A_i^r$ is deleted then we get a potential decrease of at least $\lg a_i$.
    \item If the weight of $A_i^r$ does not decrease (and $A_i^r$ was not deleted) then either $\min(A_{i+1}) = \min(B_i)= \max(B_i)$ or $\min(A_{i+1})=\max(A_{i+1})=\min(B_{i+1})$. These are the only two cases since if $\max(A_i) = \min(B_i)< \max(B_i)= \min(A_{i+1})$ then $A_i^r$ would have been deleted.
    In the first case either $A_{i+1}$ or $B_i$ is deleted. 
    In the second case either  $A_{i+1}$ or $B_{i+1}$ is deleted. In all cases we get a potential decrease of at least $\lg a_i$. 
\end{itemize}
Thus we get a decrease in potential of at least  $\lg a_{i-1} - \lg a''_{i-1} + \lg a_{i} - \lg a'_{i}$. The potential decrease from the deleted elements are counted only when considering the neighbouring segments, that is, only a constant number of times.
By a similar argument the decrease in potential for segment $B_i$ is at least $\lg b_{i-1} - \lg b''_{i-1} + \lg b_{i} - \lg b'_{i}$. For $2\leq i\leq k-1$ the potential decrease is therefore 
$$
\Omega\left(\sum_{i=2}^{k-1} \lg \frac{\max(a_i, a_{i-1}, b_i, b_{i-1})}{\min(a_i', a_{i-1}'', b_i',b''_{i-1})}\right)\;.
$$
The potential increase is still bounded by $\lg W_A +\lg W_B$ as previously. The cost of all the other parts of the merge operation can be bounded as before. 
Thus the amortized cost of the new merge operation is $O(\log U_C)$. This concludes the proof of Thm.~\ref{thm:main}.

\bibliography{bib}

\end{document}